\documentclass[useAMS,usenatbib,letterpaper]{mn2e}
\usepackage{graphicx}
\usepackage{amssymb}
\usepackage{color}
\usepackage{natbib}

\usepackage{verbatim}

\def \aj {AJ}
\def \mnras {MNRAS}
\def \apj {ApJ}

\def \aap {A\&A}
\def \nat {Nature}
\def \araa {ARAA}

\def \prd {PhRvD}
\def \apss {Astr. Sp. Sc.}

\title[Model-Independent Reconstruction of $H(a)$ with SNe Ia]{Model-Independent Reconstruction of the Expansion History of the
  Universe from Type Ia Supernovae }
\author[S.Benitez-Herrera et al.]
{S.~Benitez-Herrera$^{1}$\thanks{E-mail:
benitez@mpa-garching.mpg.de}, F.~R{\"o}pke$^{2,1}$, W.~Hillebrandt$^{1}$, C.~Mignone$^{3}$,
\newauthor
M.~Bartelmann$^{3}$, J.~Weller$^{4,5,6}$.
\vspace{0.5cm}
\\$^{1}$Max Planck Institute f\"ur Astrophysik, Karl-Schwarzschild-Str.~1,
D-85741 Garching, Germany
\\$^{2}$Universit{\"a}t W{\"u}rzburg, Emil-Fischer-Str.~31, D-97074 W{\"u}rzburg, Germany
\\$^{3}$Zentrum f\"ur Astronomie, ITA, Universit\"at Heidelberg, Albert-\"Uberle-Str.~2, 69120 Heidelberg, Germany
\\$^{4}$Excellence Cluster Universe, Boltzmannstr. 2, 85748 Garching, Germany
\\$^{5}$University Observatory, Ludwig-Maximillians University Munich, Scheinerstr. 1, 81679, Munich, Germany
\\$^{6}$Max Planck Institute f\"ur Extraterrestrische Physik, Giessenbachstr., 85748, Garching, Germany
}

\begin{document}

\date{Accepted 2011 August 26. Received 2011 August 22; in original form 2010 November 15}

\pagerange{\pageref{firstpage}--\pageref{lastpage}} \pubyear{2011}

\maketitle

\label{firstpage}

\begin{abstract}
  Based on the largest homogeneously reduced set of Type Ia supernova luminosity
  data currently available -- the Union2 sample -- we reconstruct the
  expansion history of the Universe in a model-independent
  approach. Our method tests the geometry of the Universe directly
  without reverting to any assumptions made on its energy content and
  thus allows us to constrain Dark Energy models in a straightforward
  way. This is demonstrated by confronting the expansion history
  reconstructed from the supernova data to predictions of several Dark
  Energy models in the framework of the $w$CDM paradigm. In
  addition, we test various non-standard cosmologies such as
  braneworlds, $f(R)$ and kinematical models. This is mainly intended
  to demonstrate the power of the method. Although statistical rigor
  is not the aim of our current study, some extreme cosmologies
  clearly disagree with the reconstructed expansion history. We note
  that the applicability of the presented method is not restricted to
  testing cosmological models.  It can be a valuable tool for pointing
  out systematic errors hidden in the supernova data and planning
  future Type Ia supernova cosmology campaigns.

\end{abstract}

\begin{keywords}
supernovae: general -- cosmology: cosmological parameters -- 
cosmology: observations -- cosmology: theory

\end{keywords}

\section{Introduction}

Type Ia supernovae (SNe~Ia) are currently the best (relative) distance
indicators out to redshifts of $z \sim 1$
\citep*{tammann1978a,colgate1979a,riess1996a,schmidt1998a,perlmutter1999a}. As
such, they have been instrumental in shaping our current picture of
the Universe \citep{leibundgut2001a, leibundgut2008a}. In particular,
the notion of an accelerated expansion rate of the Universe was established a
decade ago based on SN~Ia distance measurements
\citep{riess1998a,perlmutter1999a}.

For lack of a deeper understanding, the cause of this acceleration is
commonly parametrized in standard
$\Lambda$CDM cosmology as a Dark Energy component that currently dominates the energy contents of
the Universe \citep[see, e.g.,][for a recent review]{turner2007a}. The
most plausible explanation for it may be the vacuum energy (or
cosmological constant) with a constant equation of state $w=p/\rho = -1$
\citep[]{turner1997a}. In this case, Dark Energy would be an
elastic and smooth fluid exerting a repulsive gravity that produces
the observed accelerated expansion
\citep{ostriker1995a,liddle1996a,turner1997a}. 

So far, however, all attempts to compute the
vacuum energy have led to values that are about $55$--$120$ orders of
magnitude off the observed value
\citep[]{weinberg1989a,sahni2002a}. Other possibilities allow for a Dark Energy equation of state varying with
time and avoid the cosmic coincidence and the previously mentioned fine-tuning
problems \citep*[]{zlatev1999a,frieman2008a}.

These concepts describe the vacuum energy to be a
dynamical, evolving scalar-field slowly rolling toward its lowest
energy state \citep[]{freese1987a, frieman1995a,
  wetterich1995a,turner1997a, caldwell1998a, steinhardt1999a}. In this
context, the simplest parametrization of the Dark Energy equation of
state $w(a)$ depending on the cosmic scale factor $a$ reads
\citep{chevallier2001a,linder2003a}:

\begin{equation} 
  w(a)=w_0+w_\mathrm{a}(1-a),
\label{eq:par}
\end{equation}
where $w_0$ is the present-day value of the equation of state and $w_a$
accounts for its time-dependence.

Alternative approaches describe cosmic acceleration as a manifestation
of new gravitational physics rather than Dark Energy \citep[see,
e.g.,][]{deffayet2001a,deffayet2002a, carrol2004a,
  nojiri2006a,amendola2007a}. Instead of adding an extra term to the
energy-momentum tensor on the right side of Einstein's equations,
these modify the geometry terms on their left side in order to
reproduce the observations.

Hence, we face the situation that apart from $\Lambda$CDM a large
variety of cosmological models has been proposed to account for cosmic
acceleration. Testing their validity is an important but challenging
task. The present-day nearby Universe is fairly well-known and
therefore all models under consideration reproduce its
characteristics, or contain free parameters that can be tuned to do
so. This degeneracy is difficult to break with currently available
data.

Usually, Dark Energy models are constrained by starting out from a
Friedman cosmology in which the expansion function $H$ is
parametrized in terms of the contributions of radiation (r), matter
(m), curvature ($k$), and Dark Energy (de) to the energy
density as

\begin{eqnarray} 
  H^2(a)&=&H^2_0E^2(a)\nonumber\\
  &=&H^2_0 \left[\frac{\Omega_{\mathrm{r}0}}{a^4}+\frac{\Omega_{\mathrm{m}0}}{a^3} -
    \frac{\Omega_{k0}}{a^2}+\Omega_{\mathrm{de0}}F(a)\right],
\label{eq:dark1}
\end{eqnarray}
where $H_0$ and $\Omega_{\ldots 0}$ denote the Hubble constant and the
present-day density parameters corresponding to the different
components, respectively. A possible
time-dependence of the Dark Energy equation of state is captured by
the function $F(a)$
\begin{equation} 
  F(a)=\exp\left\{-3\int^{a}_{1} \left[1+w(a')\right] \frac{da'}{a'}\right\}.
\end{equation}
The cosmological parameters
occurring in Eq.~(\ref{eq:dark1}) are determined from fits to
observations and Dark Energy models are usually assessed by
confronting their predictions to these parameters.  The significance
of this approach is, however, limited since it automatically assumes a
Friedman model. When working with SNe~Ia this constraint is
unnecessary as they probe the geometry of the Universe directly and no
assumptions on the form of the energy-momentum tensor are required to
derive the expansion history of the Universe.

Consequently, our analysis follows a recently developed method
\citep[]{mignone2008a} to reconstruct the expansion history of the
Universe in a model-independent fashion from luminosity distance data.
The idea of a
model-independent reconstruction extracted straight from the data was
already proposed in \citet{starobinsky1998a} and reconstructions of
this kind were carried out by \citet{shafieloo2006a,shafieloo2007a}
using SN~Ia data, by \citet{fay2006a} adding constraints from
measurements of baryon acoustic oscillations (BAO) to the SN~Ia data,
and by \citet{daly2003a,daly2004a} combining SNe Ia
luminosity distances with angular-diameter distances from radio
galaxies. \citet*{seikel2008a,seikel2009a} tested the significance of
cosmic expansion directly from SN~Ia data in a model-independent way.

The goal of this work is to apply the method of \citet{mignone2008a}
to the most complete SN~Ia data set currently available and to
constrain some flavors of Dark Energy models. This is intended as a
\emph{demonstration} of the general power of SN~Ia distance
measurements for testing Dark Energy models when analyzed in a
model-independent way. A rigorous statistical treatment that would
let us rule out specific models is beyond the scope of this
paper.

The paper is organized as follows. In Section~\ref{sec:method}, the
essential aspects of the model-independent methodology are
reviewed. The application of the method to luminosity-distance
measurements is discussed in Section~\ref{sec:data}. A comparison
between SN~Ia data and several Dark Energy models is presented in
Section~\ref{sec:application}. In Section~\ref{sec:alternative}, the
predictions for the expansion history of the Universe of several
non-standard cosmologies are confronted with our reconstruction from
SN~Ia data.  Ways of improving the reconstruction with new data that
may become available through current surveys are pointed out in
Section~\ref{sec:improve}. This shows that our model-independent
analysis can be instrumental in planning observational
campaigns. Moreover, it can also point to potential systematics in the
data that do not generally arise from traditional ways of
analysis. This aspect is highlighted in
Section~\ref{sec:systematics}. Finally, conclusions are drawn and
future perspectives are discussed.

\section{Model-independent method}
\label{sec:method}

The minimal assumptions adopted by \citet{mignone2008a}
are that the
expansion rate is a reasonably smooth function and that the Universe is topologically
simply connected, homogeneous and isotropic, i.e.\ it
is characterized by a Robertson-Walker metric:

\begin{equation}
  ds^2=c^2dt-a^2(t)\left[d\chi^2+f^2_k(\chi)(d\theta^2+\sin^2\theta d\phi^2)\right]
\end{equation}

The Robertson-Walker metric allows us to define an
angular-diameter distance by
\begin{equation} 
  D_{\mathrm{A}}(a)=a f_k[\chi(a)],
\end{equation}
with the comoving angular-diameter distance 
\begin{equation} 
  f_k(\chi)=\left\{
\begin{array}{cl}
  \sin\chi & (k=1{\rm ,~~spherical)}\\
  \chi & (k=0{\rm ,~~flat)}\\
  \sinh\chi & (k=-1{\rm ,~~hyperbolic)}
\end{array} \right.
\label{eq:comoving}
\end{equation}
and the comoving distance
\begin{equation} 
  \chi(a)=\frac{c}{H_0}\int^{1}_{a}\frac{dx}{x^2E(x)}.
\end{equation}

Through Etherington's relation \citep{etherington1933a}, which holds
for any space-time, we can relate the luminosity distance to the
angular-diameter distance:
\begin{equation}
  D_{\mathrm{L}}(a)=\frac{1}{a^2}D_{\mathrm{A}}(a),
\end{equation}
This allows us to write the former as an integral of the inverse of
the expansion rate
\begin{equation} 
  D_{\mathrm{L}}(a)=\frac{c}{H_{0}}\frac{1}{a}\int^{1}_{a} 
  \frac{dx}{x^2E(x)}\equiv\frac{c}{H_{0}}
  \frac{1}{a}\int^{1}_{a}\frac{dx}{x^2}e(x),        
\label{eq:lum_dist}                
\end{equation}
with $e(a)\equiv E^{-1}(a)$ and $k=0$ set in Eq.~(\ref{eq:comoving})
for simplicity of notation. It is worth noticing that the choice of
$k=0$ does not affect the fundamental method and can be dropped
without change of principle if needed.  The Hubble length $c/H_0$, by
which the luminosity distance is scaled, shall be dropped in the
following for the sake of brevity.

From Eq.~(\ref{eq:lum_dist}) the derivative with respect
to $a$ is taken
\begin{equation} 
  D'_{\mathrm{L}}(a)=-\frac{1}{a^2}\int^{1}_{a}\frac{dx}{x^2}e(x)-\frac{e(a)}{a^3}.                           
\end{equation}
This expression can be transformed to a Volterra integral equation of
the second kind for the unknown function $e(a)$,
\begin{equation} 
  e(a)=-a^3D'_{\mathrm{L}}(a)+\lambda\int^{a}_{1}\frac{dx}{x^2}e(x),
\label{eq:volterra}                        
\end{equation}
with the inhomogeneity $f(a)=-a^{3}D'_{\mathrm{L}}(a)$ and the simple
kernel $K(a,x)=x^{-2}$. The general parameter $\lambda$ will later be
fixed to $\lambda=a$. For now it is introduced to make the connection
to a class of equations for which solutions are known to exist and to
be uniquely described in terms of a Neumann series
\citep[see][]{arfken1995a}:
\begin{equation}
  e(a)=\sum^{\infty}_{i=0}\lambda^ie_{i}(a).
\end{equation}
A possible choice for the function $e_i$ would be
\begin{eqnarray}
  e_0(a)&=&f(a),\\
  e_{n}(a)&=&\int^a_1K(a,t)e_{n-1}(t)dt,
\end{eqnarray}
where, for the guess of $e_0$, the approximation of either the integral
or $\lambda$ to be small has been made in Eq.(\ref{eq:volterra}). This
approximation is valid in all important cosmological cases and it is
subsequently improved until convergence is reached. It essentially means that starting, for instance, 
from the $\Lambda$CDM cosmology observations say that deviations must be small, 
if they exist at all.


Equation (\ref{eq:volterra}) involves the derivative of the luminosity
distance with respect to the scale factor $a$.  Observations of SNe~Ia
provide measurements of the distance modulus, $\mu_i$, and redshifts,
$z_i$ (or scale factors $a_i=(1+z_i)^{-1}$). Thus one could think of
taking the derivative of the luminosity distance directly from the
data. However, this is an inconvenient procedure since the result
would be extremely noisy and the determination of $D'_{\mathrm{L}}(a)$
would be unreliable.  Therefore, it is necessary to suitably smooth
the data in the first place by fitting an adequate function
$D_{\mathrm{L}}(a)$ to the measurements $D_{\mathrm{L}}(a_i)$.  The
derivative in Eq.~(\ref{eq:volterra}) can then be approximated by the
derivative of $D_{\mathrm{L}}(a)$.  Thus, the derivative of the fitted
data is taken as a representation of the derivative of the real data.
The method proposed by \citet{mignone2008a} achieves this goal via the
expansion of the luminosity distance $D_{\mathrm{L}}(a)$ into series
of (in principle) arbitrarily chosen orthonormal functions $p_j(a)$:
\begin{equation}
  D_{\mathrm{L}}(a)=\sum^{J}_{j=0}c_jp_j(a).
\label{eq:DL_fit}
\end{equation}
Suitable orthonormal function sets can be constructed by Gram-Schmidt
orthonormalisation from any linearly independent function set. The $J$
coefficients $c_j$ are those which minimize the $\chi^2$ statistic
function when fitting to the data.  Therefore, with this
representation of the data, the derivative of the luminosity distance
function is simply given by
\begin{equation}
  D'_{\mathrm{L}}(a)=\sum^{J}_{j=0}c_jp'_j(a).
\end{equation}
Due to the linearity of Eq.~(\ref{eq:volterra}), it is possible to
solve it for each mode $j$ of the orthonormal function set
separately. Thus, the final solution in terms of a Neumann series is
\begin{equation}
  e(a)=\sum^{J}_{j=0}c_je^{(j)}(a),
\end{equation}
\label{eq:final_ex}
that is, the measured coefficients of the series expansion give the
solution for the expansion function. 

The number of terms to be included in order to produce an acceptable
fit to the data depends on the choice of the orthonormal basis.  Due
to the limited quality of current observational data, however, it is
expected that the coefficients of this expansion can only be
determined up to a certain order. Therefore, although the basis of the
expansion is arbitrary with ideal data, it is not in practice. A
preferred choice would minimize the number of required
coefficients. \citet{maturi2009a} suggested an optimal basis system
derived from principal component decomposition of cosmological
observables. The use of this optimal basis would certainly reduce the
coefficients needed to obtain an accurate reconstruction and we shall
do so in future work. In the current analysis, however, we settle for
the arbitrarily chosen basis proposed by \citet{mignone2008a}, which
uses the linearly independent set
\begin{equation}
  u_j(x)=x^{j/2-1},
\label{eq:basis}
\end{equation}
orthonormalized via the Gram-Schmidt method.  The orthonormalization
interval is chosen to be $[a_\mathrm{min},1]$, where
$a_\mathrm{min}=(1+z_\mathrm{max})^{-1}$ is the scale factor
corresponding to the maximum redshift $z_\mathrm{max}$ in the
supernova sample. This way, an arbitrary set of orthonormal functions
${p_j(a)}$ is constructed. When projecting the luminosity distance
$D_{\mathrm{L}}(a)$ onto these basis functions, we can solve for the
expansion coefficients.  In this case, at least the first five
coefficients are different from zero
\citep[see][]{mignone2008a}. However, as shown in Table~\ref{table},
only the first three can be determined with current data; the fourth
and fifth coefficients lose significance. With future space-based
telescopes, new and better-quality data will be become available for
this kind of analysis. This should procure an improvement on the
accuracy of the reconstruction by allowing for determining more
coefficients in the expansion.

\section{Application to SN~Ia data: the Union2 sample}
\label{sec:data}

We apply the model-independent method to the largest homogeneously
reduced SN~Ia sample currently available. The Union2 sample
\citep{amanullah2010a} consists of 557 SNe~Ia. It includes the
recently extended dataset of distant supernovae observed with the HST
\citep{riess2007a,amanullah2010a}, the data from the SNLS
\citep{astier2006a}, ESSENCE \citep{miknaitis2006a,wood-vasey2007a}
and SDSS \citep{holtzman2008a}
surveys, several compilations from literature
\citep[e.g.][]{hamuy1996a}, and the new data from nearby
SNe~Ia of \citet[]{hicken2009a}.

\begin{figure}
 \includegraphics[keepaspectratio=true,width=91mm,clip=true]{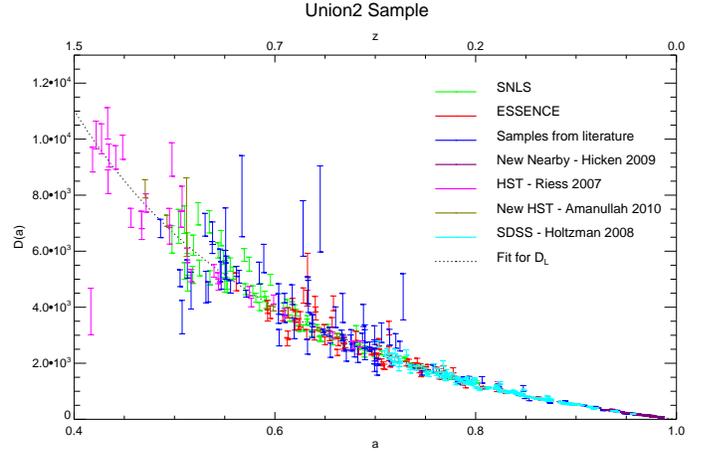}
 \caption{Luminosity distance plotted against scale factor $a$ for the
   the Union2 sample. The dotted line is the best fit for the
   luminosity distance obtained by fitting the data to Eq.~(\ref{eq:DL_fit}) and using three expansion coefficients.}
\label{fig:dl}
\end{figure}

In Fig.~\ref{fig:dl} the Union2 sample is shown, together with the
best fit of the luminosity distances to Eq.~(\ref{eq:DL_fit}) when
using the first three terms in the expansion (dotted line, see
Table~\ref{table} for values). There is considerable scatter around the fit, mainly
introduced by the distant SNe which are more difficult to calibrate
and are usually afflicted by host-galaxy extinction and survey
selection effects.

In the following two sections, we compare the expansion history
derived from the Union2 sample with predictions of various
cosmological models. This comparison, however, is not fully
consistent. SNe~Ia are not standard candles but have to be calibrated
as distance indicators applying an empirical calibration.  This
calibration procedure introduces a dependence between the calibrated
measurements that gives rise to finite non-diagonal entries in the
covariance matrix for the distance modulus.

Moreover, the parameters
of the calibration are determined simultaneously with the cosmological
parameters  in the SALT2 light curver fitter \citep{guy2007a} assuming a $\Lambda$CDM
model. As the Union2 sample was calibrated with SALT2, this applies to
the data used in the following and implies that the parameters are
model-dependent. This clearly contradicts
the intended model-independence of our method and would prevent
precise fits to non-$\Lambda$CDM cosmologies. Such fits, however, are
not our goal here. Apart from that, $\Lambda$CDM happens to be an
extraordinary good fit to the data and models with more degrees of
freedom do not improve the quality of the fit more than statistically
expected. That means that the correction parameters will not change
significantly if one assumes a different model, as long as it roughly
matches $\Lambda$CDM. Models that show a $H(z)$ drastically different
from $\Lambda$CDM could also have different parameters in the
calibration, but are not interesting as they would not agree with the
data. Additionally, the calibration parameters should be redshift
independent and therefore, essentially decoupled from
cosmology. Consequently, we argue that the induced covariance is small
and the variation of the parameters with cosmology is negligible for
our purposes.

\section{Comparison to $\Lambda$CDM and Dark Energy models}
\label{sec:application}

\subsection{$\Lambda$CDM cosmology}
\label{sect:lcdm}

\begin{figure}
\includegraphics[keepaspectratio=true,width=91mm,clip=true]{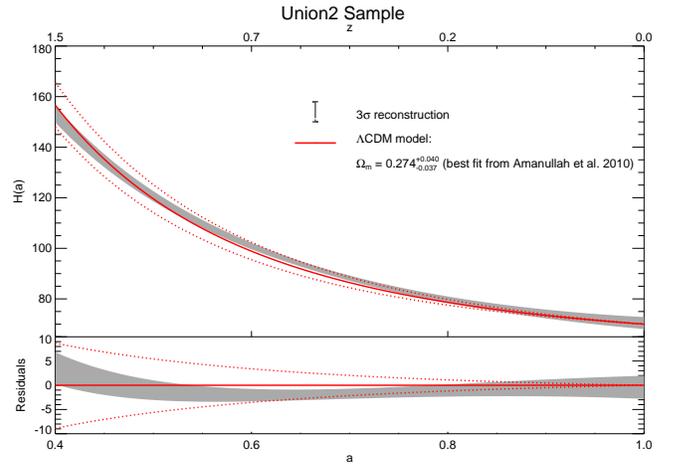}
\caption{The reconstructed expansion rate, with $3\sigma$ errors,
  extracted from the Union2 sample. The red lines represent a $\Lambda$CDM model
  assuming the best-fit cosmological parameters reported by
  \citet{amanullah2010a} . It is shown here only for the sake of comparison, this model was not used in our calculations. The bottom panel shows the residuals between
  the reconstruction and the model. The dotted lines are the limits of
  the model varying $\Omega_\mathrm{m}$ with $H_0$ fixed to $70$ km s$^{-1}$ Mpc$^{-1}$.
\label{fig:const}}
\end{figure}

The result of our reconstruction of the expansion
rate $H(a)$ from the Union2 sample is compared with a $\Lambda$CDM
model in Fig.~\ref{fig:const}. Here, we adopt the values for the
today's density parameters from the best fits to the Union2 sample
given by \citet{amanullah2010a} -- $\Omega_{\mathrm{m}} =
0.274^{+0.040}_{-0.037}$. The Hubble constant is assumed to be
$h=0.7$ ($H_0=100h$ km s$^{-1}$ Mpc$^{-1}$) and the equation of state
to be constant ($w=-1$).

Thanks to the improved quality and size of the Union2 data set, the
reconstruction has smaller errors bars than that of
\citet{mignone2008a}, who made use of the SNLS data only. The
$\Lambda$CDM model is in good agreement with the supernova data,
although its slope is slightly different. This leads to a deviation at
intermediate values of $a$, but, within the error bars, the
$\Lambda$CDM model is still consistent with our reconstruction.

\subsection{Dark Energy models}

\begin{table}
\begin{center}
\caption{The first five coefficients of the reconstructed expansion.}
 \begin{tabular}{@{}lcccccc}
Coefficients & Fitted Values  \\ 
\hline
\vspace{0.1cm}
~~~~~~c$_0$ & 2772 $\pm$ 14 &  \\ 
\vspace{0.1cm}
~~~~~~c$_1$ & -1389 $\pm$ 15 & \\
 \vspace{0.1cm}
~~~~~~c$_2$ &  73 $\pm$ 11 &  \\ 
\vspace{0.1cm}
~~~~~~c$_3$ & 12$\pm$ 5 &   \\ 
\vspace{0.1cm}
~~~~~~c$_4$ & -3 $\pm$ 1 & \\  
\end{tabular}
\label{table}
\end{center}
\end{table}

Relaxing the assumption of constant equation of state, we
analyze a Dark Energy model proposed by \citet*{rapetti2005a}. 
It is an extension of the parametrization in Eq.~(\ref{eq:par})
proposed by \citet{chevallier2001a} and \citet{linder2003a}, which assumes a fixed transition redshift ($z_\mathrm{t}=1$) between the
current value of the equation of state and the value at early times,
$w_\mathrm{et}=w_0+w_1$. In contrast, the model discussed by
\citet{rapetti2005a} introduces $z_\mathrm{t}$ as an extra free
parameter so that the equation of state $w$ can be written as

\begin{equation}
  w(a)=\frac{w_\mathrm{et}z+w_0z_\mathrm{t}}{z+z_\mathrm{t}}=\frac{w_\mathrm{et}(1-a)a_\mathrm{t}+w_0(1-a_\mathrm{t})a}{a(1-2a_\mathrm{t})+a_\mathrm{t}},
\end{equation} 
where $a_\mathrm{t}$ is the transition scale factor.  From the Friedmann
equation, the expansion function in terms of redshift is then given by

\begin{equation}
  H(z)=H_0\sqrt{\Omega_{\mathrm{m}}(1+z)^3+\Omega_{\mathrm{de}}f(z)+\Omega_k(1+z)^2},
\end{equation}
with
\begin{equation}
  f(z)=(1+z)^{3(1+w_\mathrm{et})}\mathrm{e}^{-3(w_\mathrm{et}-w_0)g(z;z_\mathrm{t})};
\end{equation}
Here, the function $g(z;z_\mathrm{t})$ is defined as 
\begin{equation}
  g(z;z_\mathrm{t})=\frac{z_\mathrm{t}}{z_\mathrm{t}-1}\ln\left(\frac{z_\mathrm{t}}{z_\mathrm{t}+z}\right).
\end{equation}

\begin{figure}
\includegraphics[keepaspectratio=true,width=91mm,clip=true]{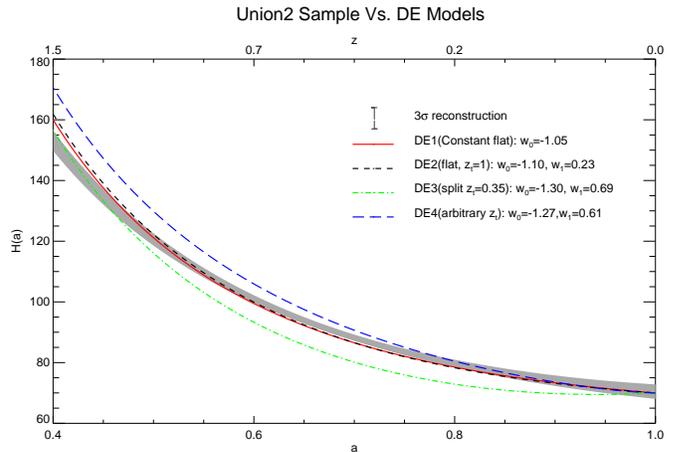}
\caption{The reconstructed expansion history, with $3\sigma$ errors, extracted from Union2
  sample and confronted with several Dark Energy
  models with the best fit cosmological parameters constrained by \citet{rapetti2005a}. }
\label{fig:d-e}
\end{figure}

\citet{rapetti2005a} constrained the best-fit cosmological parameters
for different possibilities within this model, varying the number of
free parameters. They made use of SN~Ia \citep [the Gold sample
by][]{riess2004a}, X-ray galaxy clusters and CMB data for this
analysis.  In Fig.~\ref{fig:d-e}, we compare those possibilities to
the model-independent reconstruction of the expansion function
extracted from the Union2 sample. This is meant to illustrate that
standard cosmologies within the $w$CDM paradigm with different
choices of parameters do not fit the data in all cases. This is not
easy to see in other analyses which are model-dependent.  Our
reconstruction of $H(z)$ can be a useful tool since it allows us to
rule out cosmological models based entirely on the data.  For example,
the two most extreme models shown in Fig.~\ref{fig:d-e} -- one with a
fixed $z_\mathrm{t}=0.35$ (green dashed-dotted line) which splits the SN and the
cluster data sets into similarly low and high redshift subsamples, and
the one with arbitrary $z_\mathrm{t}$ (blue dashed line) -- are clearly
inconsistent with our reconstruction. The other models (marginally) agree with the
SN measurements within the error bars. However, they have slightly
different slopes and -- as for the $\Lambda$CDM model -- some tension
exists at intermediate and small values of $a$ (intermediate and high
redshifts).

\vspace{0.3cm}
\section{Going beyond $\Lambda$CDM}
\label{sec:alternative}

Despite the good agreement with cosmological observations, the
$\Lambda$CDM model is conceptually problematic, as it leaves the
fine tuning and coincidence problems unresolved. A number of
alternatives to the $\Lambda$CDM parameterization have been suggested in
order to address those difficulties. We study some of these
alternatives here and compare them to our reconstruction.

\subsection{Modified Gravity}
\label{sec:mod}

The idea of modified gravity differs from dark energy models. It does not describe the late-time acceleration as being caused by
some unknown energy component but suggests that General Relativity may
be inaccurate at large scales. Such scenarios are likely to diverge in their predictions
for $H(z)$ from $\Lambda$CDM and therefore they can be constrained
from the reconstructed expansion history of the Universe.

\subsubsection{\bf The $f(R)$ models}

One way of modifying gravity theory is by adding terms to the Ricci
scalar ($R$) in the Einstein-Hilbert Lagrangian
\citep{starobinsky1980a,carrol2004a}.  The Palatini formalism provides
second order differential field equations that can account for the
present cosmic acceleration without any need of Dark Energy
\citep{fay2007a}. Several parameterisations of $f(R)$ are
possible. For the sake of simplicity, we follow the parametrization
of \citet{carvalho2008a} here \citep[for details on other
parameterisations see, e.g.,][]{hu2007a,fay2007a}.  The
general functional $f(R)$ is assumed to have the form

\begin{equation} 
  f(R)=R-\frac{\beta}{R^n},
\label{eq:f_R}
\end{equation} 
where $R$ is the Ricci scalar; $n$, $\beta$, and $\Omega_\mathrm{m}$, are the
parameters of the model. $\Lambda$CDM is recovered for $n=0$ The expansion rate can be written as
\begin{equation} 
\label{eq:f_R1}
  H^2(z)=H^2_0\left[\frac{3\Omega_{\mathrm{m}0}(1+z)^3+f/H^2_0}{6f'\xi^2}\right],
\end{equation}
with 
\begin{equation} 
  \xi=1+\frac{9}{2}\frac{f''}{f'}\frac{H^2_0\Omega_{\mathrm{m}0}(1+z)^3}{Rf''-f'},
\end{equation}
where the notation $f'=df/dR$, $f''=d^2f/dR^2$ has been adopted.
By evaluating these equations at $z=0$ it is possible to obtain a
relationship between $n$, $\beta$, and $\Omega_\mathrm{m}$. Thus, if
we specify the values of two of them the other is immediately
fixed. In our analysis we treat $n$ and $\Omega_m$ as free
parameters.

\begin{figure}
\includegraphics[keepaspectratio=true,width=91mm,clip=true]{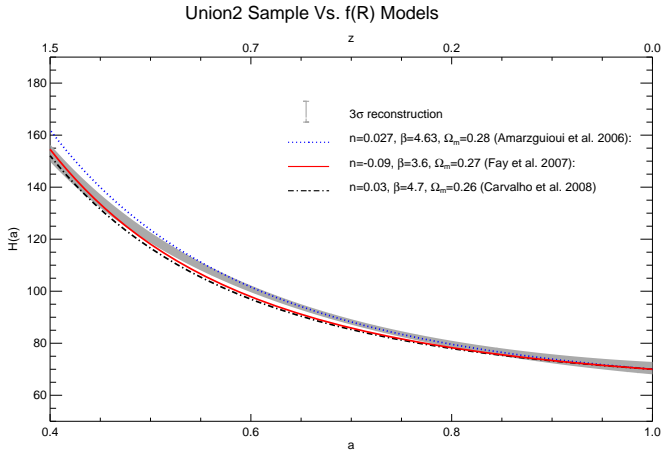}\\ 
\caption{The reconstructed expansion history, with $3\sigma$ errors, extracted from Union2 
sample compared with several $f(R)$ models. The values for the models are taken from previously 
reported fits to BAO and CMB data in combination with independent $H(z)$ (black dashed-dotted line) 
and luminosity distance measurements (red solid -- SNLS sample -- and blue dotted -- Gold sample -- lines).}
\label{fig:f_R}
\end{figure}

We compare the model-independent expansion history derived from the
Union2 sample with different $f(R)$ models in
Fig.~\ref{fig:f_R}. The adopted values for the model parameters are taken from previously reported fits
in the literature. \citet{carvalho2008a} obtained constraints on the ($n$, $\Omega_\mathrm{m}$) 
parameter space from the combination of
BAO and CMB data with independent determinations of the Hubble parameter at different redshifts. 
This model (black dashed-dotted line) has a substantially different slope and diverges from our reconstruction, 

In similar studies, other authors have combined SNe data with BAO and CMB. 
We consider here the numbers obtained by \citet{fay2007a} and \citet{amarzguioui2006a} 
using the SNLS and Gold samples, respectively. We find these models (red solid and blue dotted lines) 
closer to our reconstruction although they present deviations at high redshift (for the Gold sample) 
and at intermediate redshifts (for the SNLS sample).

\subsubsection{\bf Brane worlds}

Another possibility to modify gravity is to consider higher dimensional models.
Braneworld cosmology describes the observable Universe as a
three-dimensional brane embedded in higher dimensional spacetime
(or ``bulk''). All matter particles and fields are trapped on the brane
(i.e. conservation of mass-energy holds firm) while gravity is free to
propagate in the brane as well as in the bulk.

There are different scenarios within brane-world cosmology 
\citep[see, e.g.,][]{maartens2004a,alam2006a}. We
consider here, for example, the Dvali-Gabadadze-Porrati (DGP) model
\citep*{dvali2000a}. This model allows the extra dimension to be
large, and its generalisation to a Friedmann-Robertson-Walker brane
produces a self-accelerating solution \citep{deffayet2001a}. The
resulting Friedman equation is a modification of the General
Relativistic case and reads
\begin{equation} 
\label{eq:dgp1}
  H^2-\frac{H}{r_\mathrm{c}}=\frac{8\pi G\rho}{3},
\end{equation}
where the cross-over scale $r_\mathrm{c}$ is defined as
\begin{equation} 
\label{eq:dgp2}
  r_\mathrm{c}=\frac{1}{H_0(1-\Omega_\mathrm{m})}.
\end{equation}
The additional term $H/r_\mathrm{c}$ in Eq.~(\ref{eq:dgp1}) behaves 
like a Dark Energy component with an effective equation of state that 
evolves from $w=-1/2$ for $z \gg 1$ to $w=-1$ in the distant future.

For a $w$CDM model (assuming that Eq.~(\ref{eq:par}) holds) it 
is possible to obtain an expression for $w(a)$ which mimics the 
evolution of the DGP model \citep*[see][]{tang2006a}:
\begin{equation} 
\label{eq:dgp3a}
  w(a)=-1+\frac{\Omega_\mathrm{m}a^{-3}}{\left[(r_\mathrm{c}H_0)^{-1}+2\eta \right]\eta},
\end{equation}
with
$\eta=\sqrt{\Omega_\mathrm{m}a^{-3}+1/2\,(r_\mathrm{c}H_0)^3}$. At present-day,
this tends to
\begin{equation} 
\label{eq:dgp3}
  w(a=1)=-\frac{1}{1+\Omega_\mathrm{m}}.
\end{equation}

This modification allows for a description of the
expansion history as well as of the growth of large-scale structure. 

One can even go beyond DGP as an isolated theory and consider a
phenomenological model which is motivated by the concept of an
extra dimension with infinite extent. This is the so called mDGP model
\citep*{dvali2003a,thomas2009a}. It interpolates
between $\Lambda$CDM and the DGP model and allows for the presence of
an extra dimension through corrections to the Friedman equation by
introducing a parameter $\alpha$. The modified Friedman equation
reads
\begin{equation} 
\label{eq:mdgp1}
  H^2-\frac{H^{\alpha}}{r_\mathrm{c}^{2-\alpha}}=\frac{8\pi G\rho}{3},
\end{equation}
with the cross-over scale $r_\mathrm{c}$ defined as
\begin{equation} 
\label{eq:mdgp2}
  r_\mathrm{c}=(1-\Omega_\mathrm{m})^{\frac{1}{\alpha-2}}H^{-1}_0.
\end{equation}

\begin{figure}
\includegraphics[keepaspectratio=true,width=91mm,clip=true]{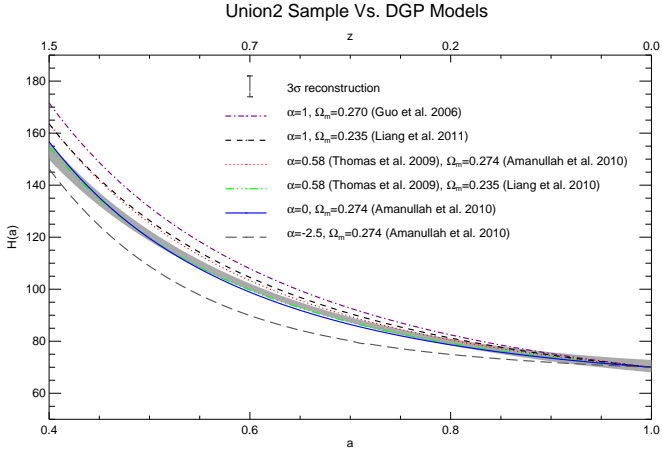}
\caption{The reconstructed expansion history, with $3\sigma$ errors, extracted from Union2 
confronted with several DGP/mDGP models. The parameters of the model are taken from previous reported fits to several cosmological data.}
\label{fig:dgp}
\end{figure}

In Fig.~ \ref{fig:dgp}, the comparison of our model-independent reconstruction 
with some DGP and mDGP models is presented.
The values of the model parameters are taken from previously reported fits 
to actual cosmological data.

We evaluate two pure DGP models ($\alpha=1$) adopting the best-fit values for $\Omega_\mathrm{m}$
reported in \citet{guo2006a} and \citet{liang2011a} (green dash-dotted and black dashed lines, respectively).
In the first case, the constraints were obtained from the Gold and SNLS samples in combination
with BAO. In the second, the best fit values are found by combining cosmology-independent 
Gamma Ray Burst and SNe~Ia data, with BAO, CMB and $H(z)$ measurements. The latter constitutes so 
far the strongest constraint obtained for the DGP model. 

We also consider the results obtained by \citet{thomas2009a} adding weak-lensing data 
to BAO and SNe (red dotted line). In that particular study, the authors find an upper 
limit for the $\alpha$ parameter ($\alpha<0.58$ at 68 per cent confidence level), 
but are not able to give constraints on the $\Omega_\mathrm{m}$ parameter. 
Therefore, and to better understand the effect of changing $\Omega_\mathrm{m}$ for a given $\alpha$, 
we make here use of the best-fit values reported in  \citet{liang2011a} 
for the DGP model and in \citet[]{amanullah2010a} for the $\Lambda$CDM model.

For the sake of comparison, a $\Lambda$CDM model corresponding to $\alpha=0$ 
(blue solid line) and a model with negative $\alpha$ (grey long-dashed line) are also included in Fig.~ \ref{fig:dgp}.

As it has been found in prior studies, \citep[see, for example,][]{maartens2006a, fairbairn2006a}, 
a pure DGP cosmology is disfavored by the SNe data. Models
with negative values of $\alpha$ also disagree with our reconstruction. 
However, it is still early to break the degeneracy betweeten
$\Lambda$CDM and mDGP models with $0\lesssim\alpha\lesssim0.50$.

\subsection{Kinematic approach}
\label{sec:kinematic}

A different way of describing the cosmic expansion history is by means
of the so-called kinematic models that do not use quantities from the
dynamic description, such as $\Omega_\mathrm{m}$ or $w$ . Here we
discuss the models of \citet{elgaroy2006a} and
\citet*{guimaraes2009a}, that are based on different parameterisations
of the deceleration parameter, $q$, and the jerk parameter, $j$.  The
deceleration parameter in terms of $z$ is defined as
\begin{equation}
  q\equiv-\frac{1}{H^2}\frac{\ddot{a}}{a}=\frac{1}{2}(1+z)\frac{(H(z)^2)'}{H(z)^2}-1, 
\end{equation}
where the primed quantity denotes the derivative with respect
to $z$. Similarly, the jerk parameter is given by
\begin{eqnarray}
  j&\equiv&-\frac{1}{H^3}\frac{1}{a}\frac{d^{3}a}{da^{3}} \nonumber \\
   &=&-\left[\frac{1}{2}(1+z)^2\frac{(H^2)''}{H^2}-(1+z)\frac{(H^2)'}{H^2}+1\right].
\end{eqnarray}

We consider five realizations here: in
the first and simplest model, $M_0$, the
deceleration parameter is constant, $q(z)=q_0$. The second model, $M_1$, is a
linear expansion of the deceleration parameter $q(z)=q_0+q_1z$. Model
$M_2$ has two phases of constant deceleration, separated by an abrupt
transition redshift, i.e.\ $q(z)=q_0$ for $z\le z_\mathrm{t}$ and
$q(z)=q_1$ for $z> z_\mathrm{t}$. The fourth model, $M_3$ is a
constant-jerk parametrization, $j(z)=j_0$. The last model, $M_4$,
derives from an expansion of the luminosity distance with free $q_0$ and
$j_0$ parameters resulting in
\begin{eqnarray}
  D_L(z)&=&\frac{c}{H_0}\left[ 
    z+\frac{1}{2}(1-q_0)z^2\right.\nonumber\\ && \left.  -\frac{1}{6}(1-q_0-3q^2_0-j_0)z^3\right].
\end{eqnarray}

All these models make specific predictions for the cosmic expansion
\citep[see][for the corresponding expressions of
$H(z)$]{guimaraes2009a}.
They are of particular interest because no assumption on the
matter-energy content present in the Universe is made. \citet{guimaraes2009a}
constrained the different parameters in their models through a Bayesian statistical analysis
from the previous Union compilation \citep{kowalski2008a}.

\begin{figure}
\includegraphics[keepaspectratio=true,width=91mm,clip=true]{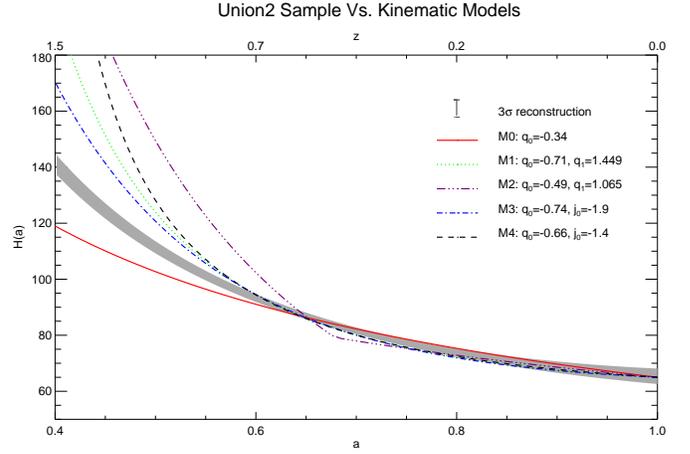}
\caption{The reconstructed expansion history, with $3\sigma$ errors, extracted from the Union2 set compared with several kinematic models with best-fit parameters obtained by \citet{guimaraes2009a}.}
\label{fig:kinetic}
\end{figure}

The comparison between the expansion history
obtained from Union2 data and the theoretical expansion function calculated 
for each model is shown in Fig.~\ref{fig:kinetic}. For the different parameters we
adopt the best-fit values obtained by
\citet[][]{guimaraes2009a}. Again, our goal here is to illustrate how the method performs and how 
families of models, with completely different parameters, can be confronted with the data in order to assess their validity.
We note that, for example, model $M_2$ could be
rejected based on current SN~Ia data because its shape differs greatly
from the reconstructed expansion history. Also $M_0$ has a different
slope which does not even fit the data at low redshifts. Models $M_1$, $M_3$
and $M_4$ agree with the SN data within the error bars up to values of
the scale factor $a\sim0.6$ ($z\sim0.66$). This is consistent with the
transition redshifts $z_\mathrm{t}$ given in \citet{guimaraes2009a}. 
The good agreement may reflect the fact that
the models were constructed in a way that mimics $\Lambda$CDM at low
redshifts. At high redshifts $z>z_\mathrm{t}$, however, none of the
models is consistent with our reconstructed expansion
history. 

\section{Improving the reconstruction with mock SN~Ia data}
\label{sec:improve}

More data at various redshifts will be obtained in future SN~Ia
surveys.  In order to test for the effects new data may have on the
reconstructed expansion function $H(z)$, we create a mock data sample
of nearby SNe~Ia with fixed $H_0$. We generate a sample of 200 mock
SNe~Ia in a $\Lambda$CDM model with $\Omega_{_\mathrm{m}}=0.274,~
\Omega_\mathrm{de}=0.726$ and $H_0=70$ km/s, with an uniform
distribution in scale factor from $0.86$ to $1$.  The reason for
adding 200 SNe is to approximately equal the number of nearby and
distant SNe contained in Union2. We calculate the luminosity distances consistent 
with equation Eq.~(\ref{eq:lum_dist}) and generate random errors with 
uncertainties 30\% smaller than those of \citet{hicken2009a}. 
The random number is drawn from a normal
distribution with mean zero and a standard deviation of one.

\begin{figure}
\includegraphics[keepaspectratio=true,width=91mm,clip=true]{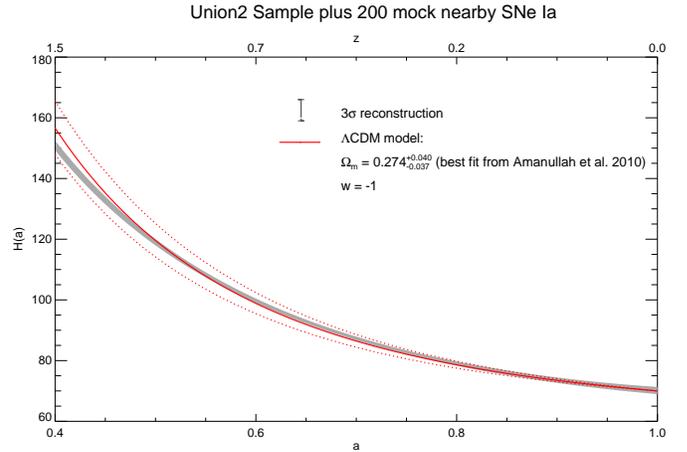}
\caption{The reconstructed expansion rate with $3\sigma$ errors
  extracted from the Union2 set, plus 200 nearby mock SNe with errors 
30\% smaller than those of \citet{hicken2009a}. Our reconstruction is compared with a $\Lambda$CDM model assuming the best-fit cosmological parameters reported by
  \citet{amanullah2010a}.}
\label{fig:u2_mock}
\end{figure}

The result of the model-independent reconstruction based on
the Union2 sample including the mock data is shown in
Fig.~\ref{fig:u2_mock}. The improvement in the reconstruction is
apparent from the significant decrease in the error bars.  The
reduction of the error at low redshifts was forced by constructing the
mock data assuming a fixed $H_0$.
This highlights the importance of a well-measured value of $H_0$ for
cosmological studies.  Although SNe~Ia are the best \emph{relative}
distance indicators out to redshifts $z\gtrsim 1$, they do not provide
reliable \emph{absolute} distances. Therefore, the zero point in our
$H(z)$ reconstruction is not well-determined. Fixing $H_0$ from other
measurements as a prior in our analysis would be
beneficial.

A $\Lambda$CDM model resulting from $\Omega_{\mathrm{m}} =
0.274^{+0.040}_{-0.037}$ \citep{amanullah2010a} is shown in
Fig.~\ref{fig:u2_mock} for comparison. This model is consistent with
our reconstruction within the error interval. However, a small
difference in slope is observed, similar to that for the original
Union2 sample (cf.\ Fig.~\ref{fig:const}). The slight deviation at high 
redshifts is probably due to the additional weight put on the mock data which have 
smaller errors.

This analysis demonstrates the power of our method
in testing redshift ranges in which more (and more accurate) SN data
would help to significantly improve the constraints on cosmological
models. Such considerations may help with the design of future surveys.

 \section{Testing for systematics in SN~Ia data compilations} 
\label{sec:systematics}

The method of model-independent reconstruction of the expansion
history of the Universe can in principle be used as a tool to study systematics in
SN~Ia data. Given, for instance, two samples from different surveys, a question to
ask is whether the respective cosmological parameters derived from
both of them predict an expansion history that is consistent with the direct
data analysis.

\begin{figure}
\includegraphics[keepaspectratio=true,width=91mm,clip=true]{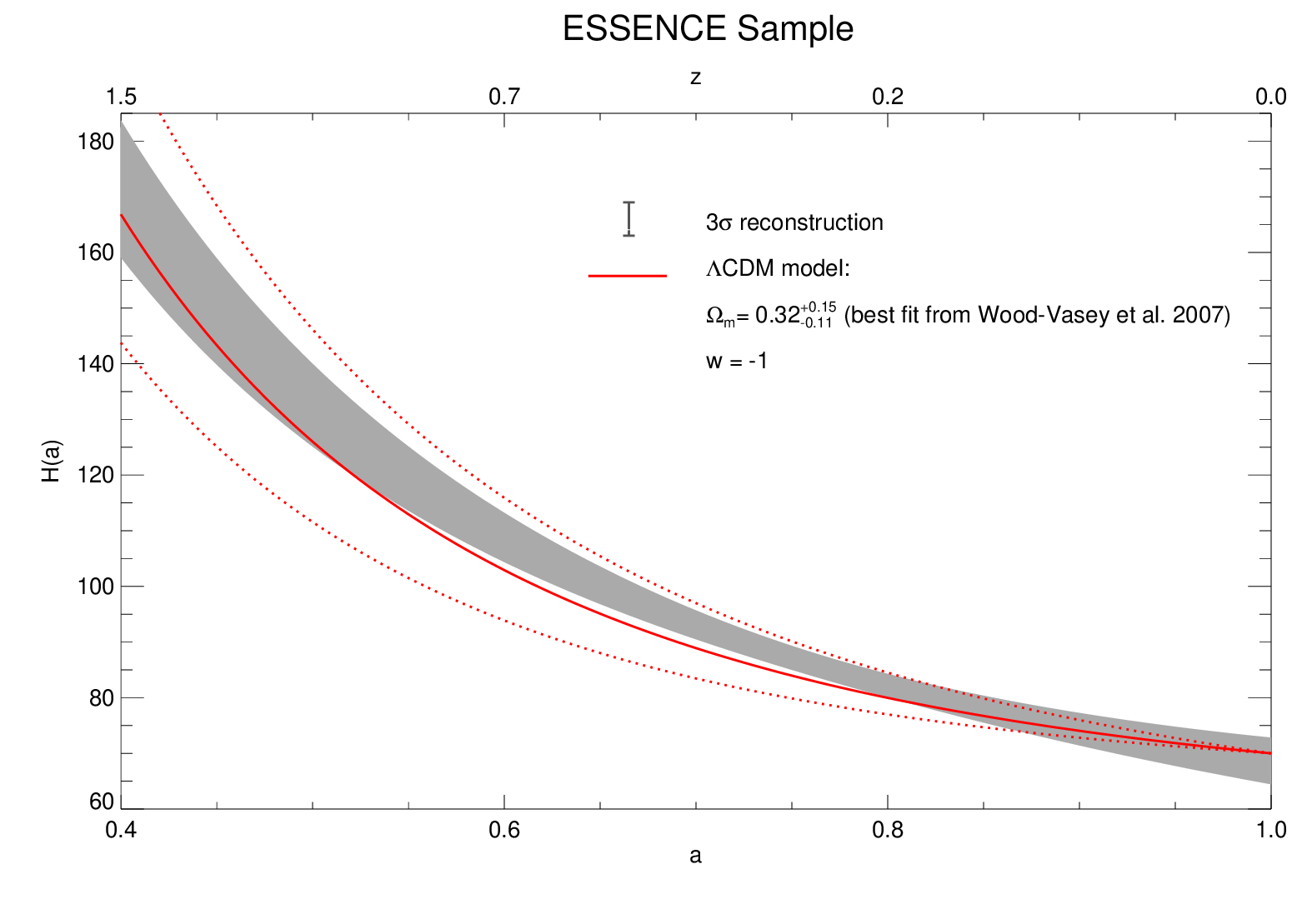}\\
\includegraphics[keepaspectratio=true,width=91mm,clip=true]{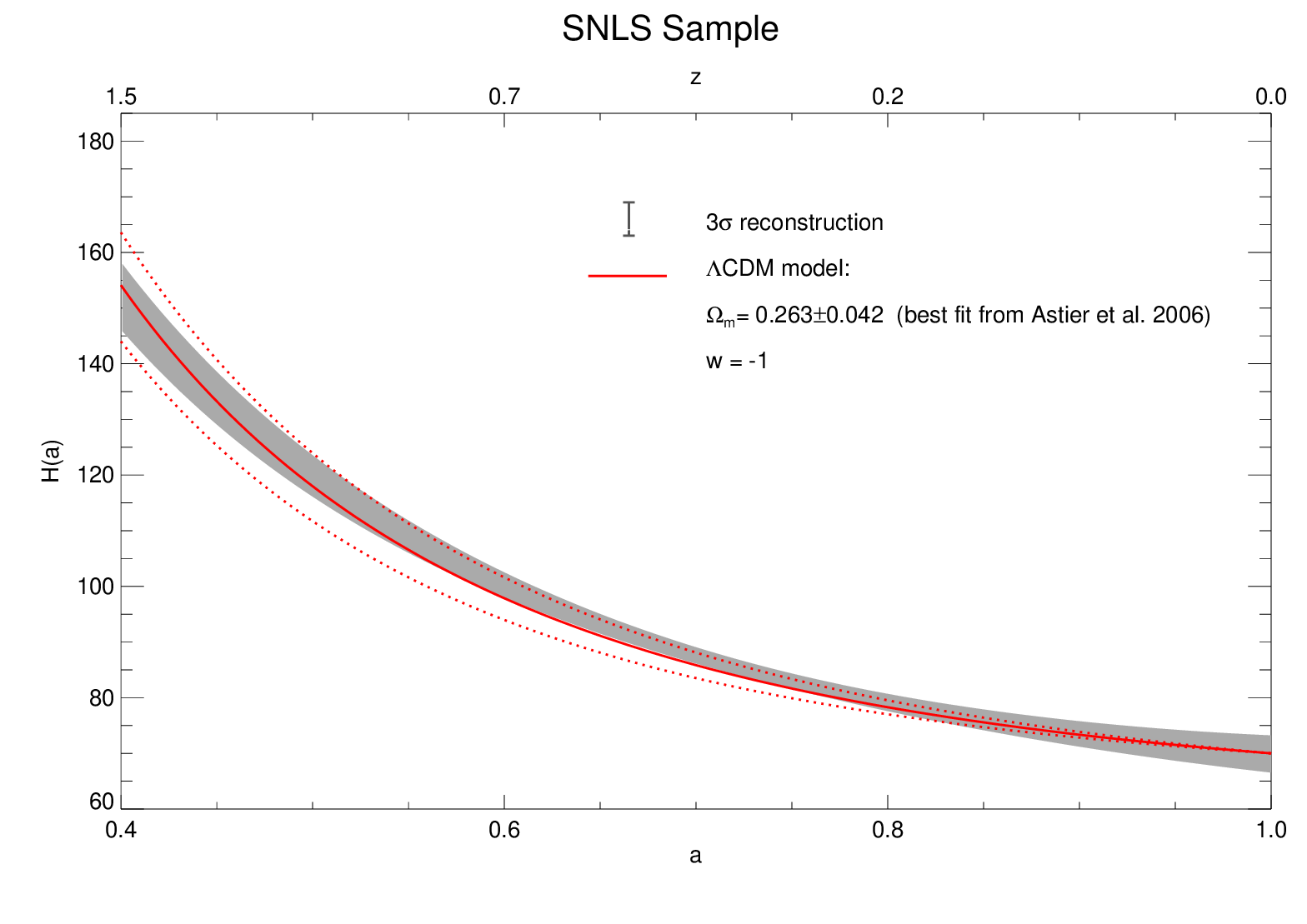}
\caption{The reconstructed expansion history with $3\sigma$ errors for the ESSENCE 
(upper panel) and SNLS (lower panel) samples compared to $\Lambda$CDM models based
 on the individual best-fit parameters. The value of $\Omega_\mathrm{m}=0.32^{+0.15}_{-0.11}$ was
obtained by \citep[]{wood-vasey2007a} combining the ESSENCE, SNLS and the gold samples \citep[]{riess2004a}.
In order to get a reliable reconstruction 57 nearby SN Ia from literature have been added to both samples.}
\label{fig:snls}
\end{figure}

We demonstrate this on the example of the Union2 data set which is a
collection of data from various SN~Ia surveys. As noted in
Sect.~\ref{sect:lcdm}, there is some tension between the reconstructed
expansion history and the predictions of a $\Lambda$CDM cosmology at
intermediate redshifts, although within the error bars they are
consistent. One possibility is that this tension could be due to
different characteristics and systematic errors of the sub-samples. In
the most extreme case, one sub-sample could be responsible for
introducing the divergence from a $\Lambda$CDM model. This can be identified by
reconstructing the expansion history for the sub-samples
individually. The two main components of the Union2 data set are the
ESSENCE and the SNLS samples, as they both cover a wide range in
redshifts with a relatively large number of objects.

The expansion histories of both samples are compared with $\Lambda$CDM
models based on the individual best-fit parameters in
Fig.~\ref{fig:snls}. Generally, the error bars for the ESSENCE data
are wider because they concentrate on a redshift range of about
$0.6<a<0.8$ ($0.25<z<0.66$), whereas the SNLS data extend up to
$a\sim0.5$ ($z\sim1$). But, apart from that, both samples show
consistency of the reconstructed expansion with a $\Lambda$CDM
model. The tension at intermediate redshifts is present in both of
them. Therefore, the deviation of the slope of the reconstructed
$H(z)$ from the $\Lambda$CDM prediction seen in the Union2 data is not
introduced by a mismatch of the main sub-samples or systematic errors
in one of them.

\section{Conclusions}

In this work, we provided further constrains on the expansion history
of the Universe in a model-independent way.  The luminosity-distance
measurements, obtained from SNe~Ia, depend only on space-time
geometry, and can be directly related to the Hubble function without
need of assuming any dynamical model.
To illustrate the power of this approach to investigate Dark Energy
models, the method was used to reconstruct the expansion history
from the Union2 sample \citep{amanullah2010a} -- the most complete
compilation of SN~Ia distances to date. Furthermore, a comparison with $\Lambda$CDM
models with different values of the matter density,
$\Omega_\mathrm{m}$, and the dark energy density, $\Omega_\mathrm{de}$,
was presented. For the Union2 sample, we found good agreement between a
$\Lambda$CDM model with the best fit parameters obtained specifically
from this sample and the data. However, the slope of the curve is not
exactly the same and a systematic deviation seems to appear at
intermediate values of the scale factor.

General relativity can accommodate the detected acceleration of the
Universe but cannot give a deeper understanding about its
cause. Non-standard cosmological models have been suggested as
alternative explanations for acceleration without dark energy. We made
use of the model-independent method to quantitatively compare the ability 
of some dark energy models and non-standard cosmologies to represent 
the data. Comparing the reconstruction recovered from the Union2 sample data with some
braneworld, $f(R)$, and kinematic models we found that none of the
considered cosmologies can be rejected rigorously on the basis of
current SN~Ia data. Particular braneworld models, such as the mDGP/DGP
with some parameters, however, as well as some $f(R)$ and some
kinematic models clearly disagree with our reconstruction and can be
ruled out. Using an optimal set of functions based on principal
component analysis instead of an arbitrary basis will strengthen this
statement.

We argue that the model-independent method is also a promising tool in
checking for biases and discrepancies in the SN~Ia cosmology
samples. This was tested by performing the reconstructions for
sub-samples of the Union2 set -- specifically the ESSENCE and the SNLS
data -- separately and comparing them with the corresponding best-fit
$\Lambda$CDM models. We find a general agreement between the different
data and the models, although some tension at intermediate redshifts
seems to be present in both. 

Although the method as employed here is indicative of viability
of dark energy models, improvements are necessary to reach full
decisive power. Apart from constructing an optimal basis for the
series expansion of the luminosity distance, the model-dependence
induced by the calibration of the SNe~Ia as distance indicators has to
be taken into account.

An increasing body of high quality data will help to reduce
systematic uncertainties and will provide a better reconstruction of
the expansion rate. To demonstrate this, we studied the effect that more
nearby SNe~Ia will have on the reconstruction of $H(a)$. We generated
a sample of 200 SNe at very low redshifts, uniformly distributed in
the range $0.86<a<1$ and with 30\% smaller errors than
those of \citet{hicken2009a}.  These mock data were added to the
Union2 sample and we found a great decrease in the size of the error
bars. This confirms the fact that not only an increased number of
distant SNe, but also higher quality data and larger number of nearby SNe
are indeed needed to reconstruct the expansion history more
accurately.

The strength of the method employed here to derive the expansion
history of the Universe is that it provides a purely geometrical
test. Contrary to many other ways of analyzing cosmological data it
does not revert to assumptions on the energy contents of the Universe
nor its dynamics.
It thus offers a complementary way of detecting possible systematic
effects which could affect the data and be overlooked within a
traditional analysis based on physically motivated
parametrization. Moreover, it can be used to discriminate
between different cosmological models and break the degeneracy in the
cosmological parameters. In future work, we will consider not only
SN~Ia data but also other cosmological probes, such as the angular
distances from BAO, in order to obtain tighter constrains on the
expansion history of the Universe. Finally, is it worthwhile noting
the potential of the method for the analysis of possible local
inhomogeneities through comparison of the expansion history in
different directions.

\section*{Acknowledgments}
It is a pleasure to thank M. Kowalski, B. Leibundgut and  R. Foley for providing helpful
comments and suggestions. We also wish to thank S. Appleby for discussions 
on the $f(R)$ and DGP models. 
S.B. acknowledges R. Angulo for helpful discussions. This work was partially 
supported by the Deutsche Forschungsgemeinschaft via the Transregional 
Collaborative Research Center ``The Dark Universe'' (TRR~33), the Emmy Noether Program (RO
3676/1-1), and the Excellence Cluster ``Origin and Structure of the
Universe'' (EXC~153).

\newpage

\bsp

\label{lastpage}

\end{document}